# Simulating Organogenesis in COMSOL: Image-based Modeling

Z. Karimaddini[1,2], E. Unal[1,2,3], D. Menshykau[1,2] and D. Iber*[1,2]
[1]Departement for Biosystems Science and Engineering, ETH Zurich, Switzerland
[2]Swiss Institute of Bioinformatics (SIB), Switzerland
[3]Developmental Genetics, Department Biomedicine, University of Basel, Switzerland
*Corresponding author: Mattenstrasse 26, CH-4058 Basel, dagmar.iber@bsse.ethz.ch

**Abstract:** Mathematical Modelling has a long history in developmental biology. Advances in experimental techniques and computational algorithms now permit the development of increasingly more realistic models of organogenesis. In particular, 3D geometries of developing organs have recently become available. In this paper, we show how to use image-based data for simulations of organogenesis in COMSOL Multiphysics. As an example, we use limb bud development, a classical model system in mouse developmental biology. We discuss how embryonic geometries with several subdomains can be read into COMSOL using the Matlab LiveLink, and how these can be used to simulate models on growing embryonic domains. The ALE method is used to solve signaling models even on strongly deforming domains.

**1. Introduction** Organogenesis is a highly dynamic process that is tightly regulated during embryogenesis. Many of the individual regulatory components, e.g. signaling molecules and their receptors, as well as their regulatory interactions have been identified in experiments. However, an integrative mechanistic understanding of the regulatory network is missing **[1]**. Mathematical modeling has a long history in developmental biology **[2,3]**. Limb development, in particular, has attracted much attention from modellers **[4]**. Early models were rather simplistic, and to this date most models are still solved on idealized domains that at most qualitatively resemble the physiological domains. However, the geometry can greatly impact the patterning process **[5]**, and it is therefore important to solve these models on physiological domains.

COMSOL Multiphysics is a versatile package that provides finite element method (FEM)-based solvers to solve a wide range of partial differential equation (PDE)-based problems on complex domains. We have used COMSOL to solve models of limb development **[5,6]**, bone development **[7]**, ovarian follicle development **[8]**, and branching morphogenesis **[9,10,11]**. In a series of papers on simulating organogenesis in COMSOL **[12,13,14,15]**, we have discussed methods to efficiently solve models for organogenesis on complex static and growing domains as well as models, which consider cells explicitly. Initially, these models were formulated on idealized geometries. Recently, we have started to take advantage of advancements in imaging techniques, which now provide us with detailed imaging data of organogenesis **[15]**. This now allows us to simulate our models on realistically growing embryonic domains in COMSOL Multiphysics **[16]**.

In this paper, we show how to use image-based data for simulations of organogenesis in COMSOL Multiphysics. As an example, we use limb bud development, a classical model system in mouse developmental biology. In the first step, computer readable geometries must be extracted from the 3D images and must then be imported into COMSOL. Many tissues contain clearly defined subdomains with different properties. These can be identified with suitable staining protocols for marker proteins or marker protein expression. We show how complex domains with subdomains can be imported. In a second step, the displacement fields between two consecutive image frames must be calculated and imported into COMSOL. Finally, the imported displacement fields can be used to simulate the domain shape evolution in COMSOL. Given the large number of stages that we use, we implement our models using Matlab LiveLink. We use the ALE method to solve our PDE-based signaling models even on strongly deforming domains. The simulation results can be compared to experimental data, and parameter values can be optimized to obtain an optimal match of model predictions and experimental results [14]. We conclude that the image-based modeling approach allows us to build realistic models of highly dynamic developmental processes, and allows us to study the combined impacts of patterning and growth.

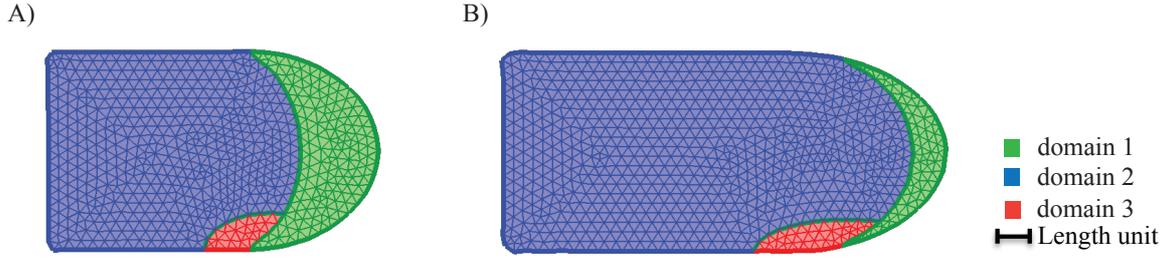

**Figure 1.** An idealized 2D limb bud domain at two different time points. The entire domain is divided into three subdomains (*domain₁* – green, *domain₂* - blue, *domain₃* - red). The domains and subdomains deform during development. (**A**) Domain at time *t*, and (**B**) at *t+1*.

## 2. Method

### 2.1 Model Formulation

Our models are defined as a set of *n* reaction-diffusion equations in the form of:

$$\frac{\partial C_i}{\partial t} + \underbrace{u\nabla C_i}_{advection} + \underbrace{C_i \nabla u}_{dilution} = \underbrace{D_i \Delta C_i}_{diffusion} + \underbrace{R_i(C_1,...,C_n)}_{reaction}$$

where $C_i$ denotes the concentration of component *i* (*n* total components), $D_i$ its diffusion constant, and $\Delta$ refers to the Laplace operator such that $D_i \Delta C_i$ describes the diffusion flux of $C_i$. In case of growing domains, advection and dilution terms have to be added to the reaction-diffusion equations; *u* represents the velocity field of the domain, for more details refer to [15]. The reaction term, $R_i(C_1,...,C_n)$, is very often non-linear and describes all reactions of component *i*, i.e. its production, degradation, and complex formation. For more details refer to [12]. The presence of species can be restricted to parts of the domain, and in that case also some reactions can become spatially restricted.

### 2.2 Regulatory Network

To illustrate our approach, we consider a concrete example. Consider a domain with 3 subdomains as shown in Figure 1, and a regulatory network that involves three components, *A*, *B* and *C* (Figure 2). All components are assumed to diffuse in the entire domains and to be degraded everywhere. Moreover, we assume that the components *A* and *C* are produced only in domain₁ and domain₃, respectively, whereas component *B* is produced in all domains. We formulate the sub domains using the unit function,

$$I_j = \begin{cases} 1 & if (x,y) \in domain_j \\ 0 & otherwise. \end{cases}$$

The spatio-temporal dynamics of the aforesaid network can be described by the following reaction terms:

$$R_A(A,B,C) = \rho_A \frac{K_{BA}^2}{K_{BA}^2 + B^2} I_{domain_1} - d_A A$$

$$R_B(A,B,C) = \rho_B (\frac{A^2}{K_{AB}^2 + A^2} \frac{K_{CB}^2}{K_{CB}^2 + C^2}) - d_B B$$

$$R_C(A,B,C) = \rho_C \frac{A^2}{K_{AC}^2 + A^2} I_{domain_3} - d_C C,$$

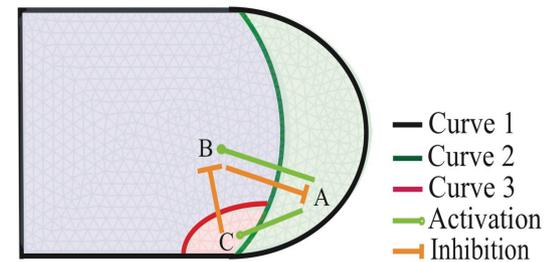

**Figure 2.** Regulatory Network. The network consists of three components, *A*, *B* and *C*. These components are produced and regulated in specific subdomains: *A* is produced and is inhibited only in domain₁ (green), *C* is activated only in domain₃ (red), and *B* is activated and repressed in the entire domain. All components diffuse in all domains.

where $\dfrac{C_j^2}{K_{C_jC_i}^2 + C_j^2}$ and $\dfrac{K_{C_jC_i}^2}{K_{C_jC_i}^2 + C_j^2}$ describe the activating and inhibitory actions of $C_j$, respectively.

*Initial and Boundary Conditions*: The initial condition of $A$ is $1.I_{domain_3}$; the initial values of $B$ and $C$ are set to zero. Zero flux boundary conditions, $\vec{n}.\nabla C_i$, are used for all components on the outer boundary, as the outer layer, the ectoderm, can be considered impermeable.

## 2.3 Boundaries and Displacement Fields

Using standard techniques for image segmentation, external and internal boundaries can be extracted [16]. This process can be repeated at different developmental time points to obtain a developmental sequence of shapes [15]. In this study, we consider the two geometries in Figure 1 as our extracted 2D geometries at two subsequent developmental time steps, $t$ and $t+1$. As can be seen, the entire domain as well as the subdomains deform from $t$ to $t+1$.

To describe the growing domains, we need to calculate the displacement fields between the two shapes at $t$ and $t+1$. A range of algorithms can be employed, which have their advantages and disadvantages dependent on the details of the geometries and their deformations (Schwaninger et al., submitted). Here, we use the uniform displacement field algorithm proposed by (Schwaninger et al., submitted): consider a curve at time $t$, $\gamma^t$, that is deformed to $\gamma^{t+1}$ within the next time step. This algorithm interpolates $N$ points on both curves:

$$\gamma^t = \{(x_1^t, y_1^t),...,(x_N^t, y_N^t)\}$$

$$\gamma^{t+1} = \{(x_1^{t+1}, y_1^{t+1}),...,(x_N^{t+1}, y_N^{t+1})\}$$

such that $\left\|(x_i^T, y_i^T),(x_j^T, y_j^T)\right\|_2$ is equal for all $i, j$ and $T \in \{t, t+1\}$. The displacement filed matrix, D, is defined as $D_i = \left[x_i^t, y_i^t, (x_i^{t+1} - x_i^t), (y_i^{t+1} - y_i^t)\right]$. For inner boundaries, it is important to use the COMSOL built-in surface-boundary parameter, $S$: every point $(x_i^t, y_i^t)$ on the curve $\gamma^t$ maps to $S_i$ and the displacement matrix is defined as $D = \left[S, (X^{t+1} - X^t), (Y^{t+1} - Y^t)\right]$. Figure 3 shows the displacement fields between two curves at two subsequent time steps $t$ and $t+1$. The displacement fields are imported into COMSOL as *Interpolation* function and are later employed in the *Moving Mesh (ale)* module to describe the domain deformation due to the growth.

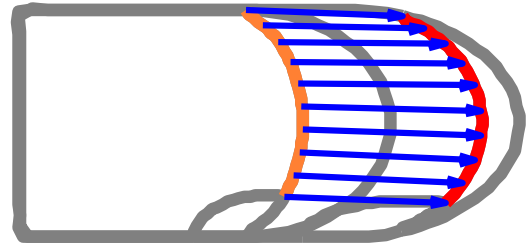

**Figure 3.** Displacement fields. Blue arrows show the displacement fields between two curves at two time steps, $t$ (orange) and $t+1$ (red).

## 2.4 Displacement of intersecting Curves

The introduction of subdomains results in intersecting boundary curves (Figure 1). The function *Interpolation Curve*, that we used to generate the boundaries, does not discriminate between intersection points and other points on the curve. All points are interpolated in the same way. Given the interpolation, there is no guarantee that the intersection point of two curves at time $t$ will be accurately displaced to their prescribed intersection point at time $t+1$. This issue can cause distorted meshes, inverted meshes, and numerical problems close to the intersection points. In case of spatially restricted variables, this inaccuracy can result in leakage of variables out of their restricted domains. From here on we will refer to the model with these problems as *Model$_1$*.

To deal with this problem, we propose the following algorithm. Assume curve$_1$ at time $t$, $\gamma_1^t$, intersects with $\gamma_2^t$ at point $P=(X,Y)$ (Figure 4A) such that

$$\gamma_1^t = \{(x_1^t, y_1^t),...,(X,Y),...,(x_n^t, y_n^t)\}$$
$$\gamma_2^t = \{(X,Y),...,(x_m^t, y_m^t)\}.$$

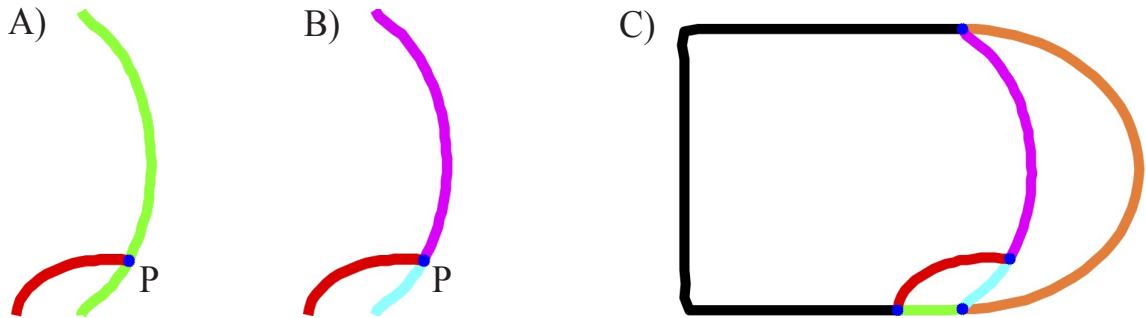

**Figure 4.** Framework to deal with intersecting boundary curves. **(A)** Curve$_1$ (green) and curve$_2$ (red) of Figure 2 intersect at point *P*. **(B)** To map the intersection point (blue point) at time *t* to the intersection point at *t+1*, the curves are divided into segments, such that the point *P* becomes the start/end point of the intersecting curves. **(C)** This algorithm is applied to all curves of Figure 2. Curve$_1$ is divided into two segments; Curve$_3$ is divided into three segments, whereas Curve$_2$ has only one segment.

Since we want to preserve the intersection point, the curves $\gamma_1^t$ and $\gamma_2^t$ have to be divided into segments such that point *P* is the start/end point of segments. We divide $\gamma_1^t$ into two segments (Figure 4B) such that:

$$\gamma_{1\ segment_1}^t = \{(x_1^t, y_1^t), \ldots, (X,Y)\}$$
$$\gamma_{1\ segment_2}^t = \{(X,Y), \ldots, (x_n^t, y_n^t)\}$$

We implemented this algorithm for all intersecting curves (Figure 4C) and determined the displacement fields for each segment individually. The coordinates of each segment and their corresponding displacement fields were then imported into COMSOL separately. Using the above algorithm, we obtain an accurate mapping of all domains and intersection points. From here on, we will refer to this model as Model$_2$.

## 3. Results

Model$_1$ and Model$_2$ were implemented in COMSOL with the parameter values as given in

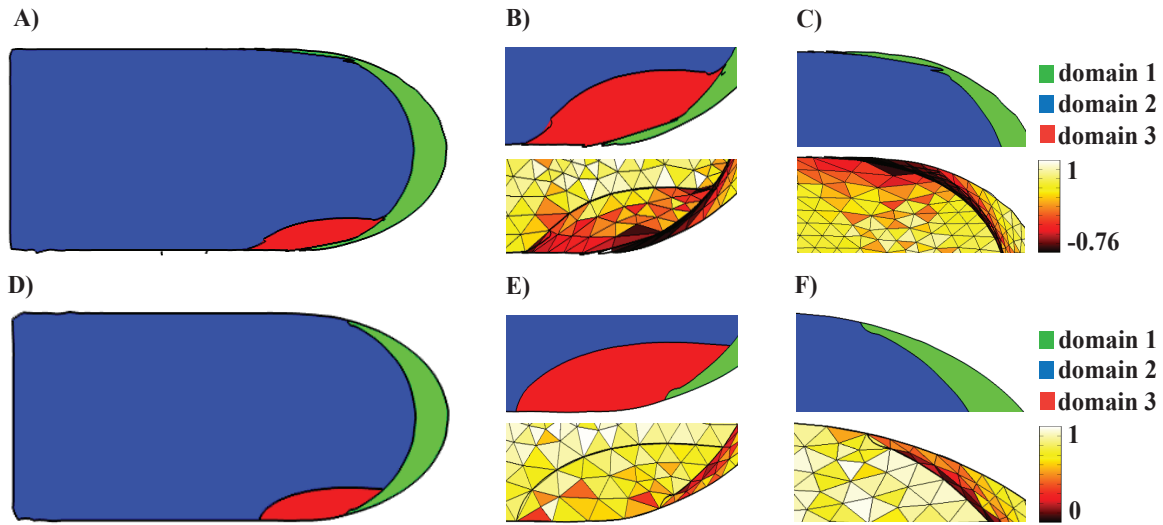

**Figure 5.** The deformed domain at the final time point. All boundary points at time *t* are mapped to their corresponding points at time *t+1* using the displacement field matrix D. **(A-C)** Final deformed domain at time *t+1* using Model$_1$. In this case, not only the intersection points are not displaced correctly, but also their adjacent points are displaced improperly. **(D-F)** Final deformed domain at time *t+1* using Model$_2$. In this case, all intersecting curves are divided into segments at their intersection points (Figure 4C). The intersection points and their neighbours are displaced perfectly. **(B,E-top)** Focus on the intersection of Curve$_1$, Curve$_2$ and Curve$_3$. **(C,F-top)** Focus on the intersection of Curve$_1$ and Curve$_2$. **(B,C-bottom)** Inaccuracies in the mapping of the intersection points result in inverted meshes close to these points in Model$_1$. **(E,F-bottom)** High quality meshes close to the intersection points in Model$_2$. The color bar indicates the quality of mesh elements; negative values indicate inverted meshes.

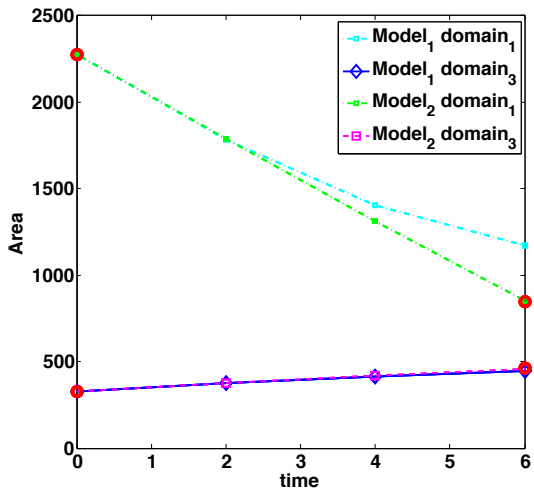

**Figure 6.** The domain areas over simulation time in $Model_1$ and $Model_2$. The real areas (red circles) of the domains at time $t$ and $t+1$ are provided for comparison.

Table 1 in the Appendix. Figure 5A-C shows that in $Model_1$ the intersection points are not displaced correctly. This problem leads to the inverted meshes close to the intersection points (Figure 5B,C bottom panels), and the subdomains are distorted (Figure 5B,C top panels) as compared to Figure 1B. Segmenting the curves at the intersection points before deformation ($Model_2$) leads to a higher quality of mesh elements and therefore accurate numerical solution and correct domain deformation (Figure 5D-F). The inverted meshes that result from inaccuracies in the mappings also lead to differences in the domain areas. Thus, Figure 6 reports the area of $domain_1$ and $domain_2$ at different time steps. As can be seen, the final area in $Model_1$ differs from the real size of the domain at $t+1$. The inverted meshes also lead to numerical problems and consequently inaccurate solutions. Thus, Figure 7 shows that the expression patterns, i.e. the effective spatial production rates, of *A* and *C* in $Model_1$ and $Model_2$ differ at time $t+1$. The components A and C have lower expression in $Model_1$ than in $Model_2$.

## 4. Conclusion

In this paper, a framework is presented to simulate PDE models on growing, embryonic domains with subdomains, using COMSOL Multiphysics. Intersecting boundaries are not mapped accurately using the standard COMSOL Interpolation function. We addressed this problem by introducing segmented boundaries.

Using this algorithm and COMSOL's Matlab LiveLink interface, one can implement also complicated domains and large sets of PDEs. This permits the simulation of large, complex regulatory networks on physiological domains. We expect that this will further increase the predictive value of the models, and will allow to better test, improve, and validate the models with experimental data.

## 5. References

1. Iber, D., Zeller, R., Making sense - data-based simulations of vertebrate limb development, *Current Opinion in Genetics & Development*, **22**, 570-577 (2012)

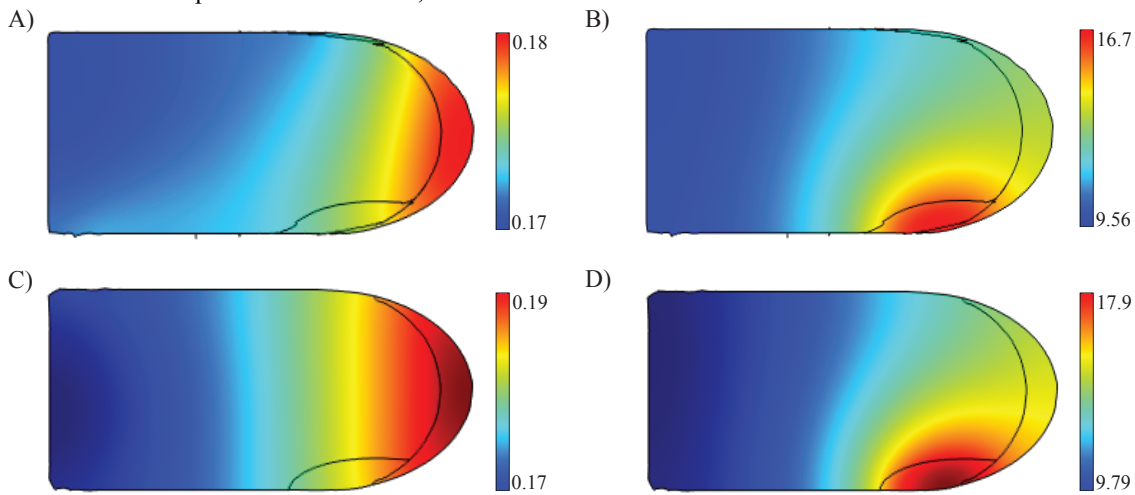

**Figure 7.** Expression patterns. Effective production rate as predicted by **(A,B)** $Model_1$ for (A) species *A*, and (B) species *C*, and by **(C,D)** $Model_2$ for (C) species *A*, and (D) species *C*.

Excerpt from the Proceedings of the 2014 COMSOL Conference in Cambridge

2. Wolpert, L., Positional information and the spatial pattern of cellular differentiation, *Journal of theoretical biology,* **25**, 1-47 (1969)

3. Turing, A. M., The chemical basis of morphogenesis. *Philosophical Transactions of the Royal Society of London. Series B, Biological Sciences*, **273**, 37-72 (1952)

4. Iber, D., Germann, P., How do digits emerge? - Mathematical Models of Limb Development, *Embryo Today, Special Issue on Musculoskeletal Development*, **102**, 1-12 (2014)

5. Badugu, A., Conradin, K., Germann, P., Menshykau, D., Iber, D., Digit patterning during limb development as a result of the BMP-receptor interaction, *Scientific Reports,* **2**, 99 (2012)

6. Lopez-Rios, J., Duchesne, A., Speziale, D., Andrey, G., Peterson, K., Germann, P., Unal, E., Liu, J., Floriot, S., Barbey, S., et al., Attenuated sensing of SHH by Ptch1 underlies evolution of bovine limbs, *Nature,* **511**, 46-51 (2014)

7. Tanaka, S., Iber, D., Inter-dependent tissue growth and Turing patterning in a model for long bone development, *Physical Biology*, **10**, 056009 (2013)

8. Bächler, M., Menshykau, D., De Geyter, C., Iber, D., Species-specific differences in follicular antral sizes result from diffusion-based limitations on the thickness of the granulosa cell layer, *Molecular Human Reproduction*, **20**, 208-22 (2014)

9. Menshykau, D., Iber, D., Kidney branching morphogenesis under the control of a ligand--receptor-based Turing mechanism, *Physical biology*, **10**, 046003 (2013)

10. Menshykau, D., Kraemer, C., Iber, D., Branch mode selection during early lung development, *PLoS computational biology,* **8**, e1002377 (2012)

11. Cellière, G., Menshykau, D., Iber, D., Simulations demonstrate a simple network to be sufficient to control branch point selection, smooth muscle and vasculature formation during lung branching morphogenesis, *Biology open*, **1**, 775-788 (2012)

12. Germann, P., Menshykau, D., Tanaka, S., Iber, D., Simulating Organogenesis in COMSOL, *Proceedings of COMSOL Conference* (2012)

13. Menshykau, D., Iber, D., Simulation Organogenesis in COMSOL: Deforming and Interacting Domains, *Proceedings of COMSOL Conference* (2012)

14. Menshykau, D., Adivarahan, S., Germann, P., Lermuzeaux, L., Iber, D., Simulating Organogenesis in COMSOL: Parameter Optimization for PDE-based models, *Proceedings of COMSOL Conference* (2013)

15. Iber, D., Tanaka, S., Fried, P., Germann, P., Menshykau, D.: Simulating Tissue Morphogenesis and Signaling. In Nelson, C., ed., *Tissue Morphogenesis: Methods and Protocols*, Springer Book Series: Methods in Molecular Biology (2013)

16. Adivarahan, S., Menshykau, D., Michos, O., Iber, D., Dynamic Image-Based Modelling of Kidney Branching Morphogenesis. In Henzinger, A., ed., *Computational Methods in Systems Biology (CMSB)*, **8130**, 106-119 (2013)

## 6. Acknowledgements

We thank the COMSOL support team, specially Sven Friedel for their support and insightful discussions. The authors acknowledge funding from the SystemsX RTD NeurostemX, a SystemsX iPhD, and a SNF Sinergia grant.

## 7. Appendix

**Table 1:** Non-dimensionalized Model Parameters with characteristic time T = 3600 sec and characteristic length L = 150 μm

| Parameter | Value | Description |
|---|---|---|
| $D$ | 1*T | Diffusion constant |
| $\rho_A$ | 1E-4*T | Production rate |
| $\rho_B$ | 50* $\rho_A$ | Production rate |
| $\rho_C$ | 200* $\rho_A$ | Production rate |
| $d_A, d_B, d_C$ | 1E-6*T | Degradation rate |
| $K_{BA}$ | 0.2 | Hill constant |
| $K_{AB}$ | 0.125 | Hill constant |
| $K_{CB}$ | 0.5 | Hill constant |
| $K_{AC}$ | 0.025 | Hill constant |